# Análisis estadístico *ex post* del conteo rápido institucional de la elección de gobernador del Estado de México en 2017

***Ex post statistical analysis of the electoral quick count for the election of the State of Mexico's governor in the year 2017***

**Arturo Erdely Ruiz***

## Resumen

En este artículo, se propone una metodología para un análisis estadístico de las estimaciones del conteo rápido institucional desde la perspectiva ideal de los resultados de los cómputos distritales de la elección de gobernador del Estado de México en 2017, particularmente en lo que atañe a aspectos como la precisión de las estimaciones, el nivel de confianza de los intervalos, el posible sesgo respecto al cómputo distrital y las conclusiones que se derivaron y reportaron. Lo anterior, con el fin de determinar el grado de cumplimiento de los objetivos de este ejercicio estadístico de carácter informativo.

* Doctor en Ciencias (Matemáticas) por la Universidad Nacional Autónoma de México (UNAM). Labora como profesor de tiempo completo en la Facultad de Estudios Superiores Acatlán de la UNAM. Pertenece al Sistema Nacional de Investigadores, nivel I. Fue miembro del Comité Técnico Asesor del Conteo Rápido (Cotecora) en la elección de gobernador del Estado de México en 2017. Sus principales líneas de investigación son análisis estadístico de datos dependientes y funciones cópula. Correo electrónico: arturo.erdely@comunidad.unam.mx

## Abstract

This paper proposes methodology to carry out a statistical analysis of the institutional electoral quick count, based on the total count of district voting in the election for the State of Mexico's governor in 2017. Its aim is to verify precision, the confidence level of interval estimations, possible bias regarding district counting, and the conclusions arrived at and reported. The main goal is establishing the degree of compliance with the objectives of the electoral quick count, which aims at providing information.

## Introducción

Por primera vez en la historia de los procesos electorales para elegir gobernador del Estado de México, en la elección de 2017 se realizó un conteo rápido institucional según lo establecido del artículo 355 al 382 del Reglamento de Elecciones del Instituto Nacional Electoral (Reine) de 2016, ejercicio que estuvo a cargo del Instituto Electoral del Estado de México (IEEM), y el que en el artículo 356 del Reine se define como sigue:

> Los conteos rápidos son el procedimiento estadístico diseñado con la finalidad de estimar con oportunidad las tendencias de los resultados finales de una elección, a partir de una muestra probabilística de resultados de actas de escrutinio y cómputo de las casillas electorales, cuyo tamaño y composición se establecen previamente, de acuerdo con un esquema de selección específico de una elección determinada, y cuyas conclusiones se presentan la noche de la jornada electoral. (Instituto Nacional Electoral [INE], 2016, p. 237)

Una vez concluida la jornada electoral, toda la información relativa al conteo rápido se hace del conocimiento público, de acuerdo con lo establecido en el artículo 382 del Reine:

> 1. A más tardar al día siguiente de la jornada electoral, y al menos durante los próximos seis meses, el Instituto y el OPL, en su ámbito de competencia, deberán publicar en sus páginas electrónicas lo siguiente:
>
>    a) El protocolo de selección de la muestra;
>    b) Las fórmulas de cálculo utilizadas para cada método establecido;
>    c) El reporte de resultados de los conteos rápidos del día de la elección, y
>    d) La base numérica utilizada en las estimaciones de los conteos rápidos, que deberá contener, al menos, la siguiente información:
>
>       I. Casillas que fueron seleccionadas en la muestra, y
>       II. Casillas que se integraron al cálculo final, cada una con el resultado de la elección.
>
> 2. Además, deberán publicar una versión de dicha información, escrita con lenguaje sencillo con el objetivo de facilitar la comprensión, entendimiento y utilidad de la realización de los conteos rápidos y sus resultados. (INE, 2016, p. 248)

Para el caso de la elección de gobernador del Estado de México en 2017, toda la información que se solicita en el artículo anterior está disponible en la página electrónica del IEEM (2017a, b y c), así como los resultados de los cómputos distritales que se obtuvieron pocos días después de la jornada electoral (véase IEEM, 2017d).

Un conteo rápido electoral es, esencialmente, un ejercicio de estimación estadística; aunque la información técnica sobre los conteos se hace pública en el caso de los institutos electorales que los llevan a cabo, en México poco se ha escrito en el ámbito científico respecto a las metodologías estadísticas utilizadas y los resultados obtenidos. Sólo hay estudios del tema acerca de la elección presidencial de 2006, en la cual, por la cerrada competencia entre los dos candidatos punteros, el entonces Instituto Federal Electoral (IFE) decidió no dar a conocer los resultados del conteo rápido, lo que dio pauta para que se publicaran algunos artículos (pocos) al respecto.

Tanto en el caso de la publicación de Eslava (2006) como en la de Mendoza y Nieto-Barajas (2016) —quienes además fueron miembros del Cotecora en las elecciones federales de 2006—, se trata de textos en que se explica más a detalle el fundamento científico de las metodologías empleadas. Aparicio (2009) realiza un análisis estadístico sobre diversos aspectos de la elección de 2006, y respecto al conteo rápido hace una defensa de la decisión del árbitro electoral de no dar a conocer su resultado la noche de la jornada, pero sin efectuar un estudio estadístico riguroso al respecto, tan sólo reiterando argumentos de Eslava (2006). Recientemente, Erdely Ruiz (2018) escribe sobre la falacia del empate técnico electoral y critica la decisión técnica de no haber dado a conocer el resultado del conteo rápido de 2006 la misma noche de la jornada. Haciendo un análisis estadístico de dichos resultados y apoyándose en Mendoza y Nieto-Barajas (2016), concluye que sí existía evidencia estadística suficiente para presentar un ganador la noche del 2 de julio.

Como se mencionó, el principal objetivo del presente trabajo es proponer una metodología estadística para analizar *ex post* las estimaciones de un conteo rápido desde la perspectiva de los resultados de los cómputos distritales, particularmente en lo que atañe a aspectos como la precisión de las estimaciones, el nivel de confianza de los intervalos, el posible sesgo respecto al cómputo distrital y las conclusiones que se derivaron y reportaron; lo dicho para poder determinar el grado de cumplimiento de los objetivos del ejercicio estadístico. Se pretende que la propuesta resulte de particular interés para especialistas estadísticos en conteos rápidos y autoridades electorales que los organizan, ya que, hasta donde fue posible revisar, no hay metodologías propuestas para este fin.

Lo anterior bien podría, además, formar parte de una propuesta metodológica de auditoría a los conteos rápidos institucionales, la que actualmente no está contemplada en el Reine, en contraste con el otro ejercicio de carácter informativo conocido como Programa de Resultados Electorales Preliminares (PREP), en el cual sí está prevista una auditoría (véase el artículo 347 del ordenamiento en cuestión).

El resultado de un conteo rápido tiene un alto impacto en el proceso electoral, ya que constituye la primera información que da a conocer un instituto comicial sobre los posibles resultados de la elección; así, aunque son estimaciones preliminares, resulta fundamental para la credibilidad del proceso que dichas estimaciones sean muy cercanas al resultado de los cómputos distritales que se dan a conocer pocos días después.

La principal pregunta de investigación del presente trabajo es la siguiente: ¿el resultado del conteo rápido en la elección de gobernador del Estado de México en 2017 es estadísticamente consistente con los resultados de los cómputos distritales que pretendió estimar? Con la metodología estadística que se propone, se contesta esta pregunta desde los principales ángulos técnicos de dicho ejercicio de estimación preliminar. Esta metodología es para aplicarse en cuanto se conozcan los cómputos distritales y así poder someter al rigor científico la efectividad técnica del conteo rápido en cuestión, y también detectar inconsistencias del mismo respecto a los cómputos distritales, para una mejor organización de futuros conteos rápidos.

## Precisión

De acuerdo con el artículo 373 del Reine:

> 1. Las muestras [de casillas electorales], entendidas como un subconjunto del espacio muestral [total de casillas], con que se inferirán los resultados de la elección respectiva, deberán cumplir con las siguientes características:
>    a) Que todas y cada una de las casillas del marco muestral construido tengan una probabilidad conocida y mayor que cero, de ser seleccionadas;
>    b) Que se utilice un procedimiento aleatorio para la selección de las muestras, que respete las probabilidades de selección determinadas por el diseño;
>    c) Que considere la posibilidad que abarque la mayor dispersión geográfica electoral posible, y
>    d) La muestra deberá diseñarse con una confianza de noventa y cinco por ciento, y con una precisión tal, que genere certidumbre estadística en el cumplimiento de los objetivos requeridos por el tipo de elección. (INE, 2016, p. 244).

El Cotecora designado por el IEEM (2017a) elaboró el documento *Criterios científicos, protocolo para selección y resguardo de la muestra para la realización del conteo rápido*, en el que se especificó que el proceso de selección de casillas para este conteo sería un muestreo aleatorio estratificado (véanse, por ejemplo, Särndal, Swensson y Wretman, 1992, p. 100; Cochran, 1980, p. 125), en el que al interior de cada estrato se seleccionarían casillas mediante muestreo aleatorio simple sin reemplazo. Como sea que se definan los estratos, éstos constituyen una partición de todo el espacio muestral (es decir, del conjunto total de casillas electorales) en subconjuntos disjuntos, de modo que, si $K_i$ es el número de casillas dentro del *i-ésimo* estrato y se eligen $c_i$ casillas, el muestreo aleatorio simple sin reemplazo dentro de dicho estrato garantiza que todas y cada una de las casillas tengan una probabilidad $\frac{c_i}{K_i} > 0$ de resultar seleccionadas dentro de la parte de la muestra que corresponda al estrato, cumpliendo con ello con los incisos a) y b) del artículo 373 del Reine. La estratificación del espacio muestral asegura que todas las casillas tengan posibilidad de ser seleccionadas en la muestra, y que de todos y cada uno de los estratos habrá contribución en la muestra que se utilice para las estimaciones del conteo rápido, conforme el inciso c) del mismo artículo.

El inciso d) del artículo 373 tiene que ver con la determinación del tamaño de muestra, esto es, de un total de $K$ casillas cuál sería un número mínimo $c < K$ de casillas tal que se cumpla el nivel de confianza requerido de 95 % y con la precisión que se establezca, que en esta elección se fijó mediante un margen de error $\varepsilon = 0.5\,\%$ (medio punto porcentual). Si $\theta$ representa el porcentaje (desconocido) de la votación que obtendrá un candidato y $\theta^*$ es un estimador puntual de la cantidad desconocida $\theta$, el cumplimiento del inciso d) del artículo 373 implica determinar un tamaño mínimo de muestra de $c$ casillas, tal que:

$$P(\,|\,\theta - \theta^*\,| \leq 0.5\,\%\,) = 95\,\% \qquad (1)$$

Si en la elección participan un total de $J \geq 2$ candidatos, entonces la estimación para cada candidato $j \in \{1, \ldots, J\}$ podría requerir un tamaño de muestra distinto $c^{(j)}$ para cumplir con (1), así que se escoge un tamaño mínimo de muestra $c$ igual al máximo de los valores $c^{(1)}, \ldots, c^{(J)}$ para poder garantizarlo. Pese a eso, el número $c$ depende de las cantidades desconocidas $\theta_1, \ldots, \theta_J$, que son justamente las que se van a estimar, por lo que el Cotecora analizó y realizó ejercicios de simulación probabilística, con distintas formas de estratificación, utilizando resultados de elecciones anteriores recientes del Estado de México (de gobernador en 2011; presidencial en 2012, en lo concerniente a la votación en la entidad, y la federal de diputados de 2015, también en lo relativo al estado).

Con base en lo anterior, dicho Cotecora estimó que era necesaria una muestra de al menos $c = 1200$ casillas de un total de $K = 18\,605$ casillas para instalarse en la elección de 2017, utilizando como estratos los 45 distritos locales del Estado de México, así como un número de casillas proporcional a la lista nominal de cada estrato. Esto es, si $n_i$ es el tamaño de la lista nominal en el *i-ésimo* estrato, entonces el número de casillas que se seleccionan por muestreo aleatorio simple dentro de dicho estrato será el número entero $c_i$ más próximo a $\frac{n_i}{n} c$, en que $n = \sum_{i=1}^{45} n_i$ es el tamaño total de la lista nominal de la elección en cuestión.

Tomando en cuenta también la experiencia del INE en las elecciones federales de 2015, en que de la muestra total solicitada, en el caso del Estado de México, sólo se recibió 66 % hacia las 10:15 p. m. del día de la jornada electoral, el Cotecora decidió pedir un número mayor de casillas (1818), ya que, de repetirse la experiencia del año mencionado, la muestra efectiva que se recibiría sería aproximadamente de 1200, que eran las casillas requeridas, lo cual era igual a 66 % de 1818. Esto último es importante enfatizarlo porque, por ejemplo, Barranco (2018), sin citar fuente alguna que le dé sustento, afirma lo siguiente: "Se prometió realizar el conteo rápido con el resultado de mil 800 casillas y finalmente no llegó a mil 300" (p. 42).

Desde un principio el Cotecora definió como tamaño de muestra requerido 1200 casillas, pero solicitó 1818, previendo que no llegara toda la muestra, de acuerdo con la experiencia anterior. Nunca se prometió realizar la estimación con 1800 casillas. Más aún, con base en el *Informe de resultados del Comité Técnico Asesor del Conteo Rápido* (IEEM, 2017b), se recibió información de 1347 casillas, un poco más de las 1200 requeridas, lo cual se confirma además con el archivo de la *Lista de casillas que se integraron en el* *cálculo final, cada una con los resultados de la votación recabada* (IEEM, 2017c), que contiene 1347 registros de casillas para la estimación del conteo rápido.

El objetivo de la presente sección es determinar, ahora con base en la información del cómputo distrital de 2017, si el tamaño de muestra de 1200 casillas estimado por el Cotecora fue el adecuado para lograr la precisión deseada con el nivel de confianza requerido de 95 %, así como la precisión alcanzable con un tamaño de muestra como el que efectivamente se recibió para el conteo rápido, que fue de 1347 casillas.

Como estimador puntual $\theta_j$ para cada $j \in \{1,\ldots,8\}$ (seis candidatos registrados, una categoría de candidatos no registrados y una categoría de voto nulo), en la fórmula (1) el Cotecora utilizó el estimador de razón siguiente:

$$\theta_j^* = \frac{\sum_{i=1}^{45} K_i y_{ij}}{\sum_{l=1}^{8} \sum_{i=1}^{45} K_i y_{il}} \qquad (2)$$

En que $y_{ij}$ es el número promedio de votos obtenido por el candidato o categoría $j \in \{1,\ldots,8\}$ en la muestra del estrato (distrito local en este caso) $i \in \{1,\ldots,45\}$. Así que, para efectos del presente análisis, mediante la misma fórmula (2) se procedió a obtener una cantidad elevada de $M$ muestras estratificadas simuladas de tamaños $c = 1200$ y $c = 1347$ a partir del cómputo distrital, y con ellas se estimó el margen de error $\varepsilon_j$ para cada candidato o categoría $j \in \{1,\ldots,8\}$ mediante el siguiente algoritmo.

Algoritmo 1:
Para cada candidato o categoría $j \in \{1,\ldots,8\}$.

1. Simular $M = 100\,000$ muestras estratificadas de $c \in \{1200, 1347\}$ casillas a partir del cómputo distrital.
2. Calcular $\theta_j^{*(1)}, \ldots, \theta_j^{*(M)}$ mediante la fórmula (2).
3. Con el porcentaje $\theta_j$ obtenido en el cómputo distrital, estimar $\varepsilon_j$ como el cuantil 95 % de $\{| \theta_j - \theta_j^{*(m)} | : m = 1,\ldots, M\}$.
4. Verificar si $\varepsilon_j \leq 0.5\,\%$.

Esta manera de estimar los valores $\varepsilon_j$ para cada candidato o categoría $j$ tiene fundamento en el modo usual de estimar cuantiles de forma no paramétrica (véanse, por ejemplo, Mood, Graybill y Boes, 1974, p. 512; Wasserman, 2006, p. 17).

Utilizando los datos del cómputo distrital, se obtiene la tabla 1, en que se aplicó el algoritmo 1 para estimar la precisión alcanzable con 1200 y 1347 casillas, respectivamente, y se agregó una última columna con las precisiones obtenidas y reportadas en las estimaciones del Cotecora. Estos resultados confirman que el tamaño y el diseño muestral eran adecuados para lograr un margen de error de no más de 0.5 % con 1200 casillas. Como finalmente se recibió un número mayor de casillas al requerido (1347), para efectos de comparación se presentan las precisiones alcanzables y alcanzadas por candidato para ese tamaño de muestra, siendo en todos los casos menores a 0.5 % y muy similares para cada candidato en particular; solamente en el caso de la categoría sobre voto nulo el margen de error del Cotecora sí resultó un tanto mayor al que debió haber sido (70 % mayor, cuando en todas las demás categorías la máxima diferencia fue de 0 a 6.7 %), aunque la razón de esto se analizará más adelante, en la sección sobre sesgo.

## Tabla 1. Precisiones alcanzables con muestras estratificadas de 1200 y 1347 casillas, respectivamente, y precisiones alcanzadas por las estimaciones del Cotecora

| Candidato | Votación alcanzada en el cómputo distrital (%) | Precisión alcanzada (%) | | Precisión alcanzada (%) por el Cotecora (1347 casillas) |
|---|---|---|---|---|
| | | 1200 casillas | 1347 casillas | |
| J. Vázquez Mota | 11.28 | 0.30 | 0.28 | 0.29 |
| Alfredo del Mazo | 33.69 | 0.47 | 0.45 | 0.42 |
| Juan Zepeda | 17.89 | 0.36 | 0.34 | 0.34 |
| Óscar González | 1.08 | 0.05 | 0.05 | 0.05 |
| Delfina Gómez | 30.91 | 0.42 | 0.39 | 0.40 |
| Teresa Castell | 2.15 | 0.06 | 0.06 | 0.06 |
| No registrado | 0.11 | 0.02 | 0.02 | 0.02 |
| Nulo | 2.89 | 0.11 | 0.10 | 0.17 |

Fuente: Elaboración propia con base en IEEM (2017b, c y d).

## Nivel de confianza

La equivalencia de la desigualdad en la ecuación (1) se expresa a continuación:

$$\theta^* - 0.5\,\% \leq \theta \leq \theta^* + 0.5\,\% \qquad (3)$$

Y nos permite la interpretación alternativa de que, con probabilidad de 95 %, el porcentaje desconocido de votos $\theta$ estará en el intervalo siguiente:

$$[\ a, b\ ] = [\ \theta^* - 0.5\,\%, \theta^* + 0.5\,\%\ ] \qquad (4)$$

En el que $\theta^*$ es un estimador puntual de $\theta$ con base en una muestra estratificada de $c = 1200$ casillas y una precisión (o margen de error) de 0.5 %. Nótese que la longitud del intervalo resultante es $b - a = 1$ %, es decir, igual al doble del margen de error de 0.5 %. Por lo tanto, el método de estimación que se utilice para el conteo rápido, en congruencia con lo anterior, tiene que calcular intervalos con un nivel de confianza de 95 %, mismos que debieran resultar con longitudes no mayores a 1 %. El nivel de confianza para los intervalos que calculó el Cotecora fue establecido en el documento *Criterios científicos, protocolo para selección y resguardo de la muestra para la realización del conteo rápido* (IEEM, 2017a), en el que se especificó lo siguiente: "todas las estimaciones del Cotecora estarán dadas en forma de intervalos de confianza del 95 %" (p. 4).

En la tabla 2 se comparan los resultados del cómputo distrital con los intervalos de confianza de 95 % estimados por el Cotecora. Como puede apreciarse, en todos los casos la longitud de los intervalos es menor a 1 %, con lo que se cumple no sobrepasando el margen de error establecido de 0.5 %.

**Tabla 2. Intervalos de confianza de 95 % estimados por el Cotecora para cada candidato o categoría, con base en la muestra estratificada recibida de 1347 casillas**

| Candidato | Intervalo del Cotecora (%) | | Longitud del intervalo (%) | Votación alcanzada en el cómputo distrital (%) | ¿Está dentro del intervalo? |
|---|---|---|---|---|---|
| | Límite inferior | Límite superior | | | |
| J. Vázquez Mota | 10.99 | 11.57 | 0.58 | 11.28 | Sí |
| Alfredo del Mazo | 32.75 | 33.59 | 0.84 | 33.69 | No |
| Juan Zepeda | 17.60 | 18.28 | 0.68 | 17.89 | Sí |
| Óscar González | 1.03 | 1.13 | 0.10 | 1.08 | Sí |
| Delfina Gómez | 30.73 | 31.53 | 0.80 | 30.91 | Sí |
| Teresa Castell | 2.15 | 2.27 | 0.12 | 2.15 | Sí |
| No registrado | 0.08 | 0.11 | 0.03 | 0.11 | Sí |
| Nulo | 2.95 | 3.29 | 0.34 | 2.89 | No |
| Participación | 53.31 | 54.15 | 0.84 | 53.74 | Sí |

Fuente: Elaboración propia con base en IEEM (2017b y d).

De los nueve intervalos estimados por el Cotecora, en dos casos (candidato Del Mazo y voto nulo) el resultado del cómputo distrital quedó fuera del intervalo correspondiente. ¿Es esto un error de estimación? No necesariamente, porque, si un intervalo se calculó a un nivel de confianza de 95 %, esto significa que, como estimador estadístico, tiene una probabilidad de 95 % de contener el valor que pretende estimar, y una probabilidad de 5 % de no hacerlo. La única forma de garantizar que todos los intervalos contendrán los valores que pretenden estimar es utilizar un nivel de confianza de 100 %, pero esto daría por resultado en todos los casos intervalos [0 %, 100 %], lo cual es tan cierto como inútil.

Con cualquier nivel de confianza que sea menor a 100 %, no es posible garantizar que todos los intervalos contengan el valor que se pretende estimar. Pero en un caso extremo tampoco resultaría aceptable que todos los intervalos fallaran en contener el valor que cada uno busca estimar, aunque estrictamente la probabilidad de que esto suceda es positiva, quizá muy pequeña, mas no de cero.

De hecho, es posible determinar las probabilidades de que cierto número de intervalos no contengan el valor que pretenden estimar. Primero se hará un cálculo aproximado con el siguiente razonamiento: se trata en este caso particular de nueve intervalos con nivel de confianza de 95 % y, por tanto, la probabilidad de que cada uno no contenga el valor que pretende estimar es de 5 %. Si se tratara de intervalos independientes entre sí (que no es estrictamente el caso, y por ello se trata de una aproximación), podría considerarse una variable aleatoria $X$ con distribución de probabilidad *binomial* con parámetros $p = 0.05$ y $r = 9$, de modo que $X$ represente el número de intervalos que no logran contener el valor que pretenden estimar, por lo cual los valores posibles para $X$ van de cero a nueve. Su función de masa de probabilidades (véase, por ejemplo, Mood, Graybill y Boes, 1974, p. 88) es la siguiente:

$$P(X = x) = \binom{9}{x}(0.05)^x(0.95)^{9-x}, \quad x \in \{0, 1, \ldots, 9\} \qquad (5)$$

Mediante la fórmula anterior, la probabilidad de que a ningún intervalo se le escape el valor que pretende estimar sería igual a $P(X = 0) = 63\,\%$, o, equivalentemente y por complemento, la probabilidad de que exista al menos un intervalo que no contenga el valor que pretende estimar es de 37 %. En el caso de la segunda tabla, ocurrió que dos intervalos no contuvieron el valor que pretendían estimar; la posibilidad de que fallen dos o más intervalos sería igual a:

$$P(X \geq 2) = \sum_{x=2}^{9} \binom{9}{x} (0.05)^{x} (0.95)^{9-x} = 7.1\ \% \qquad (6)$$

El cálculo anterior sería válido si los intervalos fuesen independientes entre sí, pero cierto grado de dependencia está garantizado, ya que la suma del porcentaje de votación obtenido por cada candidato y categoría debe ser igual a 100 %; por tanto, un cambio porcentual positivo/negativo para un candidato o categoría necesariamente implica un cambio negativo/positivo para otro(s) candidato(s) o categoría(s) para que la suma se mantenga constante en 100 %. Así que una forma de estimar las probabilidades anteriores tomando en cuenta la dependencia entre candidatos es posible mediante el siguiente algoritmo:

Algoritmo 2:

1. Simular $M = 100\ 000$ muestras estratificadas de 1347 casillas, cada una a partir del cómputo distrital.
2. Con cada muestra, estimar $\theta^*$ puntualmente para cada candidato con la fórmula (2).
3. Mediante cada $\theta^*$ del paso anterior, construir un intervalo como en (4), pero sumando y restando la precisión alcanzable para cada candidato con una muestra de 1347 casillas, de acuerdo con la tabla 1, para que en cada caso resulten intervalos de nivel de confianza de 95 %.
4. Para cada muestra y conjunto de nueve intervalos estimados, contabilizar el número $X$ de intervalos que no contuvieron el porcentaje obtenido en el cómputo distrital.

5. Calcular la proporción de muestras en las que hubo *X* intervalos que no contuvieron el porcentaje obtenido en el cómputo distrital.

En la tabla 3 se muestran los resultados de aplicar el algoritmo 2, y se comparan con las probabilidades que pueden obtenerse mediante la fórmula (5). No hay una gran diferencia porque se trata de nueve candidatos y categorías; con una cantidad menor de candidatos, la dependencia entre ellos se incrementa y dichas probabilidades comenzarían a diferenciarse más de las que se obtienen con un supuesto de independencia.

**Tabla 3. Estimación de probabilidad de que intervalos del Cotecora no contengan el porcentaje de votos del cómputo distrital, con dependencia (algoritmo 2) y con independencia (binomial)**

| Número de intervalos que no contienen el cómputo distrital | Probabilidad (%) | |
|---|---|---|
| | Algoritmo 2 | Binomial |
| 0 | 62.3 | 63 |
| 1 o más | 37.7 | 37 |
| 2 o más | 9 | 7.1 |

Fuente: Elaboración propia con base en IEEM (2017d).

Con la tabla 3 queda claro que, aun con los intervalos de confianza de 95 % ideales, la probabilidad de que los nueve intervalos contuvieran (todos y cada uno) el valor que finalmente se obtendría en el cómputo distrital era de 62.3 %, y por complemento la probabilidad de que al menos un intervalo no contuviera el porcentaje que pretendía estimar era de 37.7 %. Para el caso que nos ocupa, que de acuerdo con la tabla 2 dio como resultado que dos de los nueve intervalos estimados por el Cotecora no contuvieran el porcentaje que pretendían estimar, la probabilidad de que esto sucediera y con condiciones ideales era de 9%. Lo anterior, adaptado a una prueba de hipótesis estadística, estableciendo como hipótesis nula la siguiente:

$$H_0 : \textit{Los nueve intervalos fueron correctamente calculados}$$

El *p-valor* de la prueba sería justamente de 9%, que, con el estándar usual de utilizar 5% (o menos) como umbral para rechazar una hipótesis nula, llevaría a no rechazar $H_0$. No obstante, un *p-valor* mucho mayor quizá daría más tranquilidad que uno de 9%, el cual no se encuentra tan lejos de 5%; así que, si bien no se rechazaría la hipótesis nula, habría que ser cauto en darla por buena, por el peligro de cometer lo que se conoce en estadística como *error tipo II:* aceptar $H_0$ incorrectamente. Esto es, por sí solo, el hecho de que dos de nueve intervalos no contuvieran el porcentaje que pretendían estimar no es suficiente evidencia para descalificar la calidad de los intervalos estimados, pero tampoco es la adecuada para afirmar lo contrario. Lo dicho se clarificará más adelante, en la sección sobre sesgo.

Corona Armenta (2018), quien fue consejero electoral del IEEM durante el proceso electoral en cuestión, afirma lo siguiente:

> Tal vez la mejor evaluación del conteo rápido la hizo un asesor del consejero del INE Marco Baños, de apellido Carriedo, quien afirmó que en las elecciones del 4 de junio fallaron los conteos rápidos del Estado de México y Coahuila, que "necesitaban más solvencia en ese instrumento de información preliminar pero no la encontraron". Carriedo sostuvo que en el primer estado la falla no tuvo la dimensión de Coahuila, pero también existió. "Desestimarla con la lógica de qué tanto es tantito sería un error grave, porque en entornos crispados una décima o diez puede ser la diferencia entre confianza y sospecha". El conteo proyectó de 32.75 a 33.59% la votación para Alfredo del Mazo; pero él registró 33.69 en los cómputos [distritales]. "Se podrá decir que la tendencia de ganador fue correcta, pero el rango no fue preciso, quedó muy lejos del mínimo y rebasó el máximo previsto, otra vez". (p. 71)

Con base en lo expuesto en esta sección, descalificar una estimación por intervalo por el solo hecho de que no logró contener el porcentaje obtenido en los cómputos distritales no tiene fundamento estadístico, pues, al ser un intervalo con nivel de confianza de 95 %, existe una probabilidad de 5 % de que no lo contenga, por la variabilidad propia de la aleatoriedad inherente a un muestreo probabilístico. La realidad es que sí hay una explicación para este caso particular, misma que se aborda en la sección que sigue. Y, cuando el periodista Carriedo Téllez (2017, p. 4) hace referencia a que "el rango no fue preciso", definitivamente no habla en términos estadísticos porque la precisión tiene que ver con la longitud del intervalo, que en todos los casos estuvo dentro de la meta planteada de ser no mayor a 1 %, lo que corresponde a un margen de error de 0.5 % (véase la tabla 1).

## Sesgo

El proceso de transmisión-recepción de información de las casillas seleccionadas para la muestra del conteo rápido se realizó mediante el procedimiento que se describe a continuación, el cual se ilustra en la figura 1:

a) Capacitadores-asistentes electorales del INE (CAE) transcribieron información de las actas de escrutinio y cómputo en un formato especial para el conteo rápido.
b) Los CAE se comunicaron telefónicamente al centro de captura del IEEM para dictar la información de dicho formato.
c) Los capturistas del IEEM introdujeron la información dictada por los CAE en el sistema de cómputo ex profeso del Instituto, para ser consultada y procesada por el Cotecora.

**Figura 1. Diagrama del flujo de información desde su origen en la casilla de la muestra del conteo rápido hasta su destino para utilizarse en la estimación que realizó el Cotecora**

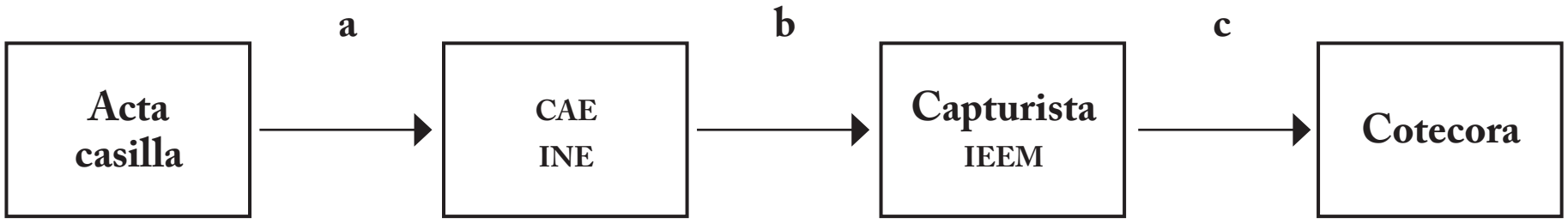

Fuente: Elaboración propia con base en IEEM (2017a).

En cada uno de los pasos (a, b y c) que aparecen en la figura 1, existe la posibilidad de que la información proveniente del acta de escrutinio y cómputo resulte alterada por la intervención humana. En principio, el supuesto es que, de haber errores, sean no intencionales y, por tanto, de naturaleza azarosa; pero, desde un rigor científico, las diferencias entre lo asentado en las actas y la información recibida por el Cotecora para realizar el conteo rápido, una vez que se conocen los cómputos distritales, son analizables estadísticamente para efecto de decidir si la hipótesis de que existe un *equilibrio razonable* entre errores de captura a favor y en contra (y, en consecuencia, errores al azar) para cada candidato es rechazable o no.

Con el objeto de dar la mayor claridad posible a lo expuesto, se presenta un ejemplo sencillo. Supongamos que se tienen datos de una moneda sobre la cual la hipótesis de interés es que se trata de una moneda equilibrada, y los datos recabados son los resultados de 21 volados que se han lanzado con ella. Si la hipótesis es correcta, se esperaría un *equilibrio* entre el número de águilas y soles obtenidos, pero ese *equilibrio* no puede tomarse de forma estricta porque, para empezar, con un número impar de volados (21 en este ejemplo), es imposible obtener el mismo número de águilas que de soles. Incluso con un número par de volados, no es difícil que con una moneda equilibrada se obtenga diferente cantidad de águilas que de soles. La ciencia estadística se encarga precisamente de cuantificar qué tan probable es que se presenten diferencias de cierta magnitud bajo la hipótesis de interés, y con base en ello decidir si es rechazable una hipótesis de moneda equi-librada, por ejemplo. Esto ilustra el enfoque que se adopta en la presente sección para el análisis de las diferencias entre la información recibida por el Cotecora respecto a las casillas de la muestra y la información de esas mismas casillas en el cómputo distrital.

En el documento en el que se presentan los resultados de los cómputos distritales de la elección en cuestión (véase IEEM, 2017d), se identificaron las 1347 casillas utilizadas por el Cotecora para el conteo rápido, y se compararon los resultados de dichas casillas con los datos de la relación de casillas que se integraron en el cálculo final (véase IEEM, 2017c). A los resultados de las 1347 casillas recibidas por el Cotecora el día de la jornada electoral se les restaron los resultados de esas mismas casillas, pero con datos del cómputo distrital posterior. Si no hubiera habido errores de captura, todas las diferencias deberían haber sido igual a cero. No fue así, y dichas divergencias son sometidas a un análisis estadístico a continuación.

Primero se presenta una estadística descriptiva de dichas diferencias, calculando para cada candidato o categoría el porcentaje de casillas en que la disparidad es cero (sin errores de captura), los porcentajes de casillas en que hubo errores de captura a favor y en contra de cada candidato, y finalmente la diferencia promedio de votos por casilla en cada caso (promedio de las diferencias) (véase la tabla 4).

## Tabla 4. Para cada candidato y categoría, porcentaje de casillas de la muestra sin error, con errores a favor y con errores en contra, y diferencia promedio de votos por casilla

| Candidato | Porcentaje sin error | Porcentaje a favor | Porcentaje en contra | Diferencia promedio |
|---|---|---|---|---|
| J. Vázquez Mota | 95.0 | 2.3 | 2.7 | 0.20 |
| Alfredo del Mazo | 84.1 | 7.6 | 8.3 | -0.19 |
| Juan Zepeda | 92.9 | 3.0 | 4.1 | 0.04 |
| Óscar González | 96.9 | 1.5 | 1.6 | -0.05 |
| Delfina Gómez | 92.6 | 3.9 | 3.5 | 0.82 |
| Teresa Castell | 95.6 | 2.5 | 1.9 | 0.06 |
| No registrado | 96.2 | 1.9 | 1.9 | -0.06 |
| Nulo | 86.8 | 8.8 | 4.4 | 0.82 |

Fuente: Elaboración propia con base en IEEM (2017c y d).

De acuerdo con la tabla 4, por ejemplo, de los candidatos registrados fue Del Mazo quien tuvo el menor porcentaje de casillas de la muestra sin error (84.1 %), con diferencias a favor en 7.6 % de ellas y en contra en 8.3 %, y la diferencia promedio de votos por casilla en su caso resultó predominantemente en contra (-0.19 votos por casilla), lo cual quiere decir que en la muestra recibida por el Cotecora este candidato estuvo subrepresentado respecto al cómputo distrital. En contraste, verbigracia, la candidata Delfina Gómez presentó 92.6 % de casillas de la muestra sin error, con 3.9 % de casillas con errores a favor y 3.5 % de casillas con errores en contra, así como una diferencia promedio de votos por casilla que resultó predominantemente a favor (+0.82 votos por casilla), lo cual significa que en la muestra recibida por el Cotecora la candidata Gómez estuvo sobrerrepresentada respecto al cómputo distrital.

Se analizará a continuación qué tanto impacto tienen la sub- y la sobrerrepresentación para cada candidato en las estimaciones del conteo rápido, y estadísticamente qué tan significativas son. En términos ideales, considerando errores humanos al azar, era de esperarse un porcentaje mayoritario de casillas sin error (como efectivamente ocurrió), un *razonable equilibrio* entre los porcentajes de casillas con errores a favor y en contra, e idealmente una diferencia promedio de votos por casilla igual a cero (véase fórmula [8] más adelante). El qué tan significativas son las desviaciones de cero de las diferencias promedio de votos por casilla para cada candidato o categoría se examinará ahora desde una óptica de estadística inferencial.

Con base en las diferencias de votos calculadas por candidato y categoría, interesa decidir para cada caso si la siguiente hipótesis estadística es rechazable o no:

$$H_0 : \textit{Las diferencias a favor y en contra ocurrieron al azar} \qquad (7)$$

En los casos en que dicha hipótesis resulte rechazable, se concluirá que existe un *sesgo* estadísticamente significativo respecto al cómputo distrital a favor o en contra, según sea de signo positivo o negativo la diferencia promedio de votos por casilla en la tabla 4.

Mediante $D$ se denotará una variable aleatoria que cuantifica la diferencia, en número de sufragios, entre la votación de un candidato en una casilla específica de la muestra recibida por el Cotecora y la votación que efectivamente se asentó en el acta de escrutinio y cómputo de acuerdo con el cómputo distrital. Se trata de una variable aleatoria discreta que asigna probabilidad positiva a que no haya error o diferencia, esto es, $P(D=0)>0$, y cuyo rango de valores satisface $\{0\}\subset Ran\,D\subset\{\ldots,-2,-1,0,+1,+2,\ldots\}$. Estrictamente, $Ran\ D$ es finito porque existe un máximo de diferencia posible, ya que la votación en cualquier casilla está acotada. Si $H_0$ es verdadera, entonces necesariamente $P(D=d)=P(D=-d)$ para todo número entero $d$, pues ante errores al azar cualquier diferencia en una casilla sería tan susceptible de resultar a favor como en contra del candidato que sea. Esto último tiene una implicación para la *esperanza* de la variable aleatoria $D$:

$$E(D) = \sum_{d \in Ran\, D} dP(D=d) = \sum_{d>0} dP(D=d) + \sum_{d<0} dP(D=d)$$
$$= \sum_{d>0} dP(D=d) - \sum_{-d>0} (-d)P(D=-d) = 0 \qquad (8)$$

Si $d_1,\ldots,d_{1347}$ son las diferencias calculadas para cada una de las casillas del conteo rápido para un determinado candidato (algunas serán cero, otras positivas, otras negativas), bajo la hipótesis de interés $H_0$, el signo positivo/negativo de las diferencias puede considerarse producto de una variable aleatoria $S$ con distribución de probabilidad uniforme sobre el conjunto $\{-1,+1\}$, esto es que $P(S=-1)=\frac{1}{2}=P(S=+1)$ (errores al azar); por tanto, el conjunto de diferencias observadas $\{d_1,\ldots,d_{1347}\}$ puede verse como una realización de la muestra aleatoria $|d_1|S_1,\ldots,|d_{1347}|S_{1347}$, en que las variables aleatorias $S_1,\ldots,S_{1347}$ son independientes e idénticamente repartidas con distribución de probabilidad uniforme sobre el conjunto $\{-1,+1\}$.

Definiendo las variables aleatorias $D_k^*=|d_k|S_k$ para $k\in\{1,\ldots,1347\}$, la diferencia promedio resultante es la variable aleatoria $D_{prom}^* = \frac{1}{1347}\sum_{k=1}^{1347} D_k^* = \sum_{k=1}^{1347} |d_k|S_k$, y, por consecuencia:

$$E(D_{prom}^*) = \frac{1}{1347}\sum_{k=1}^{1347} |d_k|\, E(S_k) =_{H_0} 0$$
$$V(D_{prom}^*) = \frac{1}{1347^2}\sum_{k=1}^{1347} |d_k|^2\, V(S_k) =_{H_0} \sum_{k=1}^{1347}\left(\frac{d_k}{1347}\right)^2 \qquad (9)$$

Lo dicho, ya que bajo $H_0$ se tiene que $E(S_k) = (-1)\frac{1}{2} + (1)\frac{1}{2} = 0$ y la varianza $V(S_k) = E(S_k^2) - [E(S_k)]^2 = (-1)^2\frac{1}{2} + (1)^2\frac{1}{2} - 0^2 = 1$. Utilizando $D_{prom}^*$ como estadístico para decidir si se rechaza o no $H_0$ dada una diferencia promedio observada para una candidato o categoría particular calculada mediante $d_{prom} = \frac{1}{1347}\sum_{k=1}^{1347} d_k$ (véase la columna sobre diferencia promedio en la tabla 4), se calcula lo siguiente:

$$p\text{-}valor = P_{H_0}\left(|D_{prom}^*| > |d_{prom}|\right) \qquad (10)$$

En la fórmula anterior, el *p-valor* es la probabilidad de rechazar erróneamente $H_0$. La práctica estadística común es decidir rechazar $H_0$ si el *p-valor* es menor a 5%, o, bajo criterios más conservadores, si es menor a 1%, que es el que se adoptará en este análisis. La probabilidad (10) se puede estimar vía simulación, generando un número elevado de muestras $\{D_1^*, \ldots, D_{1347}^*\}$ bajo la hipótesis $H_0$ y calculando con cada una de ellas el valor correspondiente a $D_{prom}^*$. Los *p-valores* calculados para cada candidato y categoría se resumen en la tabla 5.

### Tabla 5. Para cada candidato y categoría, *p-valores* de rechazar $H_0$ como se definió en (7), y diferencia promedio de votos por casilla en la muestra del conteo rápido

| Candidato | *p-valor* (%) | Dif. promedio |
|---|---|---|
| J. Vázquez Mota | 4 | 0.20 |
| Alfredo del Mazo | 51 | -0.19 |
| Juan Zepeda | 79 | 0.04 |
| Óscar González | 9 | -0.05 |
| Delfina Gómez | ‹1 | 0.82 |
| Teresa Castell | 6 | 0.06 |
| No registrado | 3 | -0.06 |
| Nulo | ‹1 | 0.82 |

Fuente: Elaboración propia con base en IEEM (2017c y d).

Lo que es posible concluir estadísticamente a partir de lo anterior es que difícilmente podrían atribuirse a errores al azar las diferencias promedio por casilla observadas para la candidata Gómez y para la categoría de voto nulo, mientras que para los demás casos no es rechazable que se deban a errores al azar. Pero, independientemente de que se rechace o no $H_0$ para cada candidato o categoría, es importante también cuantificar el efecto de estas diferencias promedio en los intervalos estimados por el Cotecora. Para ello, se procede a estimar los intervalos a los que hubiera llegado el Cotecora si no hubiesen existido estas diferencias, utilizando las mismas precisiones alcanzadas por dicho comité (véase tabla 1), pero corrigiendo el centro del intervalo mediante el estimador puntual (2) calculado sobre las casillas de la muestra con los datos del cómputo distrital. Los resultados se presentan en la tabla 6.

## Tabla 6. Estimación por intervalos del conteo rápido, con y sin errores de captura

| Candidato | Intervalos del Cotecora con errores de captura (%) | | | Votación en cómputo distrital (%) | Intervalos del Cotecora sin errores de captura | | |
|---|---|---|---|---|---|---|---|
| | ¿Dentro del intervalo? | Inferior | Superior | | Inferior | Superior | ¿Dentro del intervalo? |
| J. Vázquez Mota | Sí | 10.99 | 11.57 | 11.28 | 10.98 | 11.56 | Sí |
| Alfredo del Mazo | No | 32.75 | 33.59 | 33.69 | 32.96 | 33.80 | Sí |
| Juan Zepeda | Sí | 17.60 | 18.28 | 17.89 | 17.67 | 18.35 | Sí |
| Óscar González | Sí | 1.03 | 1.13 | 1.08 | 1.05 | 1.15 | Sí |
| Delfina Gómez | Sí | 30.73 | 31.53 | 30.91 | 30.65 | 31.45 | Sí |
| Teresa Castell | Sí | 2.15 | 2.27 | 2.15 | 2.14 | 2.26 | Sí |
| No registrado | Sí | 0.08 | 0.11 | 0.11 | 0.10 | 0.13 | Sí |
| Nulo | No | 2.95 | 3.29 | 2.89 | 2.74 | 3.08 | Sí |
| Participación | Sí | 53.31 | 54.15 | 53.74 | 53.05 | 53.89 | Sí |

Fuente: Elaboración propia con base en IEEM (2017b, c y d).

Aunque se explicó en una sección anterior que en un momento dado el resultado del cómputo distrital puede quedar fuera de algún intervalo de confianza de 95 % por el simple hecho de que existe una posibilidad de 5 % de que así suceda en un muestreo probabilístico, en este caso en particular queda claro que fueron errores de captura los causantes de que así ocurriera para el candidato Del Mazo; por tanto, es falso que esto se deba a una falta de "solvencia en ese instrumento de información preliminar" (Carriedo, 2017, p. 4). De esta forma, por ejemplo, la subrepresentación de Del Mazo en la muestra, debido a errores de captura que bien pueden considerarse al azar, provocó que el intervalo estimado por el Cotecora tuviese extremos menores a los debidos. Por su parte, la sobrerrepresentación de la candidata Gómez y del voto nulo, como resultado de errores de captura que difícilmente son atribuibles al azar, provocó que los intervalos estimados por el Cotecora tuviesen extremos mayores a los debidos.

En relación con lo anterior, sobre lo que dice Corona Armenta (2018): "no puedo afirmar que el conteo rápido del 4 de junio de 2017 se haya efectuado sin sesgo" (p. 70), es posible señalar que únicamente respecto a la votación correspondiente a la candidata Gómez y a la categoría de voto nulo en la muestra del conteo rápido se presentó un sesgo estadísticamente significativo en cuanto al cómputo distrital (a favor en ambos casos), pero que no altera el resultado en lo que atañe al lugar que ocupa cada candidato en la estimación del conteo rápido.

En el caso de Alfredo del Mazo, los errores de captura tienen más apariencia de fallos al azar, pero fueron en cantidad y variabilidad suficientes para provocar que el intervalo estimado por el Cotecora no contuviera el porcentaje obtenido por dicho candidato en el cómputo distrital. De hecho, no debiera sorprender que Del Mazo tuviera el menor porcentaje de casillas sin diferencias (84.1 %, véase tabla 4), ya que fue el único postulado por una coalición de partidos políticos: Partido Revolucionario Institucional (PRI), Partido Verde Ecologista de México (PVEM), Partido Nueva Alianza (Panal) y Partido Encuentro Social (PES). Esto implica que había un total de 15 combinaciones distintas para votar y contabilizar sufragios para di-cho candidato,[1] aumentando significativamente con ello la posibilidad de errores de captura en comparación con otros candidatos postulados por un solo partido o de forma independiente. El caso que sí resulta sorprendente es el de la categoría sobre voto nulo, con un porcentaje similar de casillas sin diferencias (86.8 %). No queda clara la razón de semejante cantidad de errores, siendo una sola categoría y con la agravante además de presentar un sesgo estadísticamente significativo respecto al cómputo distrital.

## Informe de resultados

En el artículo 380 del Reine, segundo apartado, se establece lo siguiente:

> El día de la jornada electoral, el Cotecora deberá rendir un informe de avance de la integración de la muestra al Consejo General u Órgano Superior de Dirección respectivo. El informe deberá realizarse cada hora *a partir de las 21:00 horas* de ese día y hasta la entrega de los resultados finales que haga el Cotecora a los propios Consejos Generales. (INE, 2016, p. 247; cursivas nuestras)

En el documento *Criterios científicos, protocolo para selección y resguardo de la muestra para la realización del conteo rápido*, el Cotecora estimó que "se esperaría contar con un tamaño de muestra efectivo de $n = 1200$ casillas alrededor de las 10:00 p. m. del 4 de junio [de 2017]" (IEEM, 2017a, p. 15).

De acuerdo con el *Informe de resultados del Comité Técnico Asesor del Conteo Rápido* (IEEM, 2017b, p. 1), a las 21:10 horas el Cotecora había recibido una muestra de 1347 casillas con información suficiente de los 45 estratos (distritos electorales locales) considerados para conformar el diseño muestral, y hacia las 21:15 horas procedió a elaborar el informe de resultados.

Lo anterior deja claro que el Cotecora cumplió con el tiempo marcado por el Reine y que, incluso, obtuvo una muestra ligeramente mayor a las 1200 casillas que previamente había considerado suficientes para realizar las estimaciones de acuerdo con los niveles de precisión y de confianza establecidos. Esto es importante enfatizarlo porque Barranco (2018) pregunta lo siguiente: "¿Por qué no se respetó lo anunciado y se adelantó la difusión del resultado del conteo rápido?" (p. 42). El autor mencionado no especifica qué se anunció ni por quién, mucho menos aporta una fuente al respecto, pero no hubo adelanto alguno, ya que el informe se realizó dentro de los tiempos que marca el Reine y de conformidad con los criterios científicos previamente establecidos.

---

1 Combinaciones posibles: 1. PRI, 2. PVEM, 3. Panal, 4. PES, 5. PRI-PVEM, 6. PRI-Panal, 7. PRI-PES, 8. PVEM-Panal, 9. PVEM-PES, 10. Panal-PES, 11. PRI-PVEM-Panal, 12. PRI-PVEM-PES, 13. PRI-Panal-PES, 14. PVEM-Panal-PES y 15. PRI-PVEM-Panal-PES.

En el tercer apartado del artículo 380 del Reine se establece lo siguiente:

> Sea cual fuere la muestra recabada y los resultados obtenidos, el Cotecora deberá presentar un reporte al Consejo General u Órgano Superior de Dirección que corresponda, en el que indique, además, las condiciones bajo las cuales se obtuvieron los resultados, *así como las conclusiones que de ellos puedan derivarse*. Las estimaciones deberán presentarse en forma de intervalos de confianza para cada contendiente. (INE, 2016, p. 247; cursivas nuestras)

El informe de resultados antes mencionado contiene tres apartados. En el primero, "Información general de la muestra recuperada", se cumple con reportar las condiciones en las cuales se obtuvieron los resultados. En el segundo, "Estimación de tendencias de la votación", se reportan estimaciones en forma de intervalos de confianza para cada contendiente. En el tercer apartado, "Conclusiones", se presentan, como el nombre lo señala, las conclusiones que de los resultaron pudieron derivarse, y cuya redacción fue la siguiente:

> A partir de la muestra recuperada y de acuerdo con las estimaciones realizadas por el Comité Técnico Asesor del Conteo Rápido, existe una diferencia estadísticamente significativa entre el primer y segundo lugar de la elección para gobernador del Estado de México. (IEEM, 2017b, p. 2)

Barranco (2018) afirma que "la fatídica noche del 4 de junio, ante resultados tan apretados, el presidente del Consejo del IEEM —Pedro Zamu-

dio— se atrevió a declarar que 'la diferencia [era] significativa' a favor de Del Mazo" (p. 42).

La realidad es que el único "atrevimiento" del presidente del IEEM fue leer textualmente las conclusiones del informe de resultados redactado por el Cotecora. El objetivo en la presente sección es analizar *ex post* si dicha conclusión es estadísticamente justificable desde la perspectiva del cómputo distrital. Para ello, se procedió a simular una gran cantidad de muestras estratificadas (100 000) de tamaño 1347 a partir del listado de casillas del cómputo distrital, y con cada una de ellas se estimó de forma simultánea $(\theta_1^*, \theta_2^*)$, lo cual, para el caso de la elección que se estudia, corresponde a las estimaciones puntuales por medio de la fórmula (2) de los candidatos Del Mazo y Gómez (en ese orden), y se calculó la diferencia $\theta_1^* - \theta_2^*$. El histograma correspondiente a dichas divergencias es el que se presenta en la figura 2.

A las diferencias mencionadas se aplicó la prueba de normalidad de Shapiro-Wilk (1965), y con un *p-valor* de 41 % no se rechaza la hipótesis de que las divergencias tengan una distribución *normal*, con media de 2.78 % (que coincide con la diferencia en el cómputo distrital entre el porcentaje obtenido por Del Mazo y el conseguido por Gómez, esto es 33.69 % - 30.91 % = 2.78 %, como era de esperarse) y desviación estándar de 0.38 %. La gráfica de una función de densidad de probabilidad *normal* con dichos parámetros se exhibe también en la figura 2, y la posibilidad de que se presentara una diferencia negativa (y, por consiguiente, que se concluyera que Gómez era quien triunfaba) con dicha distribución de probabilidad resulta tan pequeña como 0.000000000011 %. Por lo tanto, *ex post* la conclusión del Cotecora tiene un riesgo mínimo de ser errónea, así que resulta estadísticamente justificable.

**Figura 2. Histograma de 100 000 diferencias calculadas como porcentaje de votos a favor de Del Mazo menos porcentaje de votos a favor de Gómez en muestras estratificadas de 1347 casillas a partir del cómputo distrital**

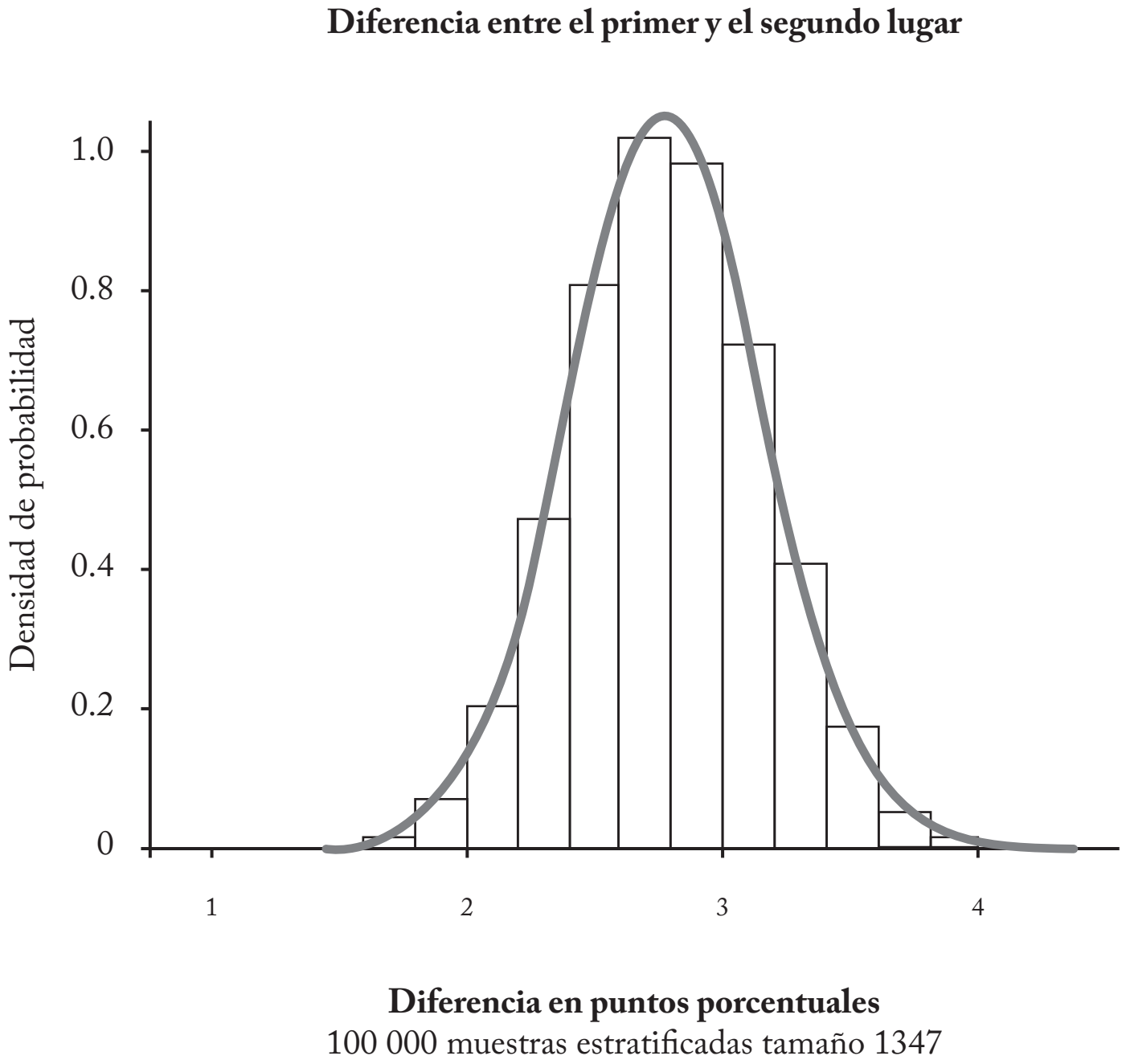

Fuente: Elaboración propia con base en IEEM (2017d).

## Conclusiones

La principal aportación de este trabajo es una propuesta de metodología estadística para analizar retrospectivamente las estimaciones de un conteo rápido electoral, misma que fue aplicada al caso del conteo rápido realizado por el Instituto Electoral del Estado de México para la elección de gobernador en 2017, con el fin de evaluar el grado de cumplimiento de los objetivos de este ejercicio estadístico de carácter informativo. Se pretende que esta propuesta sea de particular interés y utilidad tanto para especialistas estadísticos en conteos rápidos institucionales como para las autoridades electorales que los realizan.

Lo anterior es relevante por dos razones principales: primero, por tratarse del primer ejercicio de este tipo en la historia de las elecciones para gobernador del Estado de México; segundo, porque, desde que se hacen conteos rápidos institucionales en México, a la fecha no se analiza ni se audita si estadísticamente dichos ejercicios cumplen técnicamente con sus objetivos, a pesar de que tienen un alto impacto, ya que es la primera información que da a conocer un instituto electoral sobre los resultados de una elección. Incluso, esta propuesta metodológica bien podría ser aplicada a los resultados de los simulacros previos al día de los comicios, para poder prevenir potenciales fuentes de sesgo en la información que recibirá el Cotecora la noche de la jornada electoral.

La metodología propuesta demostró que el tamaño de muestra efectivo de 1200 casillas al que aspiraba el Cotecora sí cumplía con lograr la precisión planeada[2] en las estimaciones por intervalos con nivel de confianza de 95 %. Más aún, la solicitud de una muestra total de 1818 casillas, tomando en consideración una experiencia previa que daba cuenta de que no llega toda la información que se pide, resultó adecuada, pues finalmente se tuvo una muestra efectiva de 1347 casillas para realizar la estimación del conteo rápido, poco más de lo estrictamente requerido. Esto refuta señalamientos que al respecto hace Barranco (2018) sobre el tamaño de muestra utilizado por el Cotecora para efectuar las estimaciones.

La calidad estadística de una estimación por intervalo no puede evaluarse de forma binaria considerándose exitosa si logra contener el valor que pretende estimar y como un fracaso si no, como lo hace Carriedo Téllez (2017), por ejemplo. Lo expuesto porque para cada intervalo, al ser calculado con un nivel de confianza de 95 %, existe una probabilidad de 5 % de no contener el valor que pretende estimar, incluso bajo condiciones ideales.

En el caso que se estudia, estimar nueve intervalos con nivel de confianza de 95 % no se traducía en la misma probabilidad de que todos contuvie-ran el valor que pretendían estimar, pues en este caso resultaba de 62.3% (véase la tabla 3). De los nueve intervalos reportados por el Cotecora, fueron dos los que no contuvieron el valor que pretendían estimar (véase la tabla 2), y, aun bajo condiciones ideales, la probabilidad de que dos o más intervalos no incluyeran dicho valor era de 9%.

---

2 Con un margen de error de cuando más 0.5 %; véase la tabla 1.

En la primera figura se aprecian los tres momentos en que el flujo de información de las actas de escrutinio y cómputo de las casillas de la muestra para el conteo rápido puede resultar alterado por la intervención humana. Lo anterior fue causante de diferencias entre la información que recibió el Cotecora para realizar las estimaciones por intervalo la misma noche de la jornada electoral y la que debió haber recibido y quedó expresada en el cómputo distrital días después. Estas diferencias fueron analizadas estadísticamente buscando determinar para cada candidato si las discrepancias en la información podrían considerarse errores al azar. La conclusión resultante es que así fue, excepto en dos casos (candidata Gómez y categoría voto nulo), en los cuales se detecta un sesgo estadísticamente significativo (a favor, en ambos) en la información recibida por el Cotecora respecto a la información del cómputo distrital.

El caso del candidato Del Mazo es de señalarse porque, si bien las divergencias lucen como errores al azar, la variabilidad fue tal que la diferencia promedio de votos por casilla en contra (-0.19) fue suficiente para que el intervalo estimado por el Cotecora no capturara el porcentaje de votación que finalmente dicho candidato obtendría en el cómputo distrital. Aquí es de destacarse que, como se dijo anteriormente, Del Mazo fue el único postulado por una coalición de cuatro partidos políticos y, por tanto, había un total de 15 combinaciones distintas para votar y contabilizar votos para este candidato, lo que aumentaba significativamente la posibilidad de errores de captura en comparación con los otros contendientes.

De no haber existido diferencias entre la información que se recibió la noche de la jornada electoral y la que quedó asentada en actas de acuerdo con el cómputo distrital días después, todos los intervalos estimados por el Cotecora habrían contenido los porcentajes que del cómputo distrital se pretendían estimar (véase la tabla 6). A pesar de lo anterior, las diferencias no resultan de magnitud suficiente como para que el Cotecora hubiese concluido algo distinto a lo que reportó la misma noche de la jornada. La lección por aprender en este caso es la importancia de tener mejores controles en el flujo de información (véase la figura 1) que eviten diferencias estadísticamente significativas entre los datos que recibe el Cotecora y los que se asientan en las actas de las casillas que quedan en la muestra del conteo rápido; es por ello que la metodología que se propone debiera aplicarse incluso en los simulacros de conteo rápido previos al día de la jornada electoral.

Finalmente, respecto al informe de resultados realizado por el Cotecora la noche de la jornada electoral, además de cumplir con su entrega en tiempo y forma, contiene una conclusión sobre una diferencia estadísticamente significativa entre el primer y el segundo lugar de la contienda (Del Mazo sobre Gómez), la cual, a la luz del cómputo distrital y con un tamaño de muestra como el que se recibió (1347 casillas), resulta estadísticamente justificable, ya que tenía una probabilidad de error muy pequeña (aproximadamente de 0.000000000011%).

A pesar de algunas inconsistencias detectadas, con la metodología estadística propuesta puede concluirse que, en términos generales, el conteo rápido institucional de la elección de gobernador del Estado de México de 2017 cumplió adecuadamente con sus objetivos estadísticos.

## Fuentes de consulta